\documentclass[journal,onecolumn]{IEEEtran}
\usepackage{amsmath}
\usepackage{booktabs}
\usepackage{latexsym}
\usepackage{mathrsfs}
\usepackage{amsthm}
\usepackage{amssymb}
\usepackage{amsfonts}
\usepackage{amsbsy}

\usepackage[shortlabels]{enumitem}
\usepackage{url}
\usepackage{array}
\usepackage{pdflscape}
\usepackage{xcolor}
\usepackage{stmaryrd}
\usepackage{verbatim}
\usepackage{braket}
\newcommand{\Ff}{{\mathbb F}}

\newcommand\cc{{\mathcal C}}        %

\def\Tr{\operatorname{Tr}}

\theoremstyle{plain}
\newtheorem{thm}{Theorem}
\newtheorem{lem}[thm]{Lemma}

\newtheorem{example}{Example}
\newtheorem{remark}{Remark}
\newtheorem{conjecture}{Conjecture}
\def\Tr{\operatorname{Tr}}

\usepackage[colorlinks=true]{hyperref}

\begin{document}
\title{Griesmer Bound and Constructions of Linear Codes in $b$-Symbol Metric}

\author{\centerline{Gaojun Luo, Martianus Frederic Ezerman, Cem G\"{u}neri, San Ling, and Ferruh \"{O}zbudak}
\thanks{G. Luo, M. F. Ezerman, and S. Ling are with the School of Physical and Mathematical Sciences, Nanyang Technological University, 21 Nanyang Link, Singapore 637371, e-mails: $\{\rm gaojun.luo, fredezerman, lingsan\}$@ntu.edu.sg.}
\thanks{Cem G\"{u}neri and Ferruh \"{O}zbudak are with the Faculty of Engineering and Natural Sciences, Sabanc{\i} University, 34956, Tuzla, Istanbul, Turkey. e-mails: $\{\rm cem.guneri, ferruh.ozbudak\}$@sabanciuniv.edu.}
\thanks{G. Luo, M. F. Ezerman, and S. Ling are supported by Nanyang Technological University Research Grant No. 04INS000047C230GRT01.}
}



\maketitle

\begin{abstract}
The $b$-symbol metric is a generalization of the Hamming metric. Linear codes, in the $b$-symbol metric, have been used in the read channel whose outputs consist of $b$ consecutive symbols. The Griesmer bound outperforms the Singleton bound for $\Ff_q$-linear codes in the Hamming metric, when $q$ is fixed and the length is large enough. This scenario is also applicable in the $b$-symbol metric. Shi, Zhu, and Helleseth recently made a conjecture on cyclic codes in the $b$-symbol metric. In this paper, we present the $b$-symbol Griesmer bound for linear codes by concatenating linear codes and simplex codes. Based on cyclic codes and extended cyclic codes, we propose two families of distance-optimal linear codes with respect to the $b$-symbol Griesmer bound.
\end{abstract}


\section{Introduction}\label{sec:intro}
A conventional model for information transmission across noisy channels partitions a message into \emph{discrete information units}. The writing and reading processes operate on symbols that are clearly separated. Emerging storage technologies, however, have a scenario where the write resolution is significantly higher than the read resolution. Consequently, writing and reading individual symbols in consistent manners are infeasible in channels where the outputs comprise overlapping pairs of symbols. To surmount the physical constraints inherent in such channels, Casutto and Blaum in \cite{Cassuto2011}, followed by Casutto and Litsyn in \cite{Cassuto2011a}, introduced and presented early constructions of codes in the $2$-symbol metric. Within this novel framework, the outputs and errors no longer manifest as individual symbols, but rather as {\it overlapping pairs of adjacent symbols}. Yaakobi, Bruck, and Siegel generalized the earlier results to codes in the $b$-symbol metric in \cite{Yaakobi2016}.

\subsection{Linear Codes in the $b$-Symbol Metric}

Let $q = p^r$ be a prime power, with $p$ being a prime and $r$ a positive integer. Let $\Ff_q$ be the finite field with $q$ elements and let $\Ff_q^*=\Ff_q\setminus \{0\}$. We denote by $\Ff_q^n$ the $n$-dimensional space over $\Ff_q$. An $[n,k,d_{\rm H}]_q$ linear code $\cc$ is a $k$-dimensional subspace of $\Ff_q^n$ with minimum Hamming distance $d_{\rm H}$. The {\it $b$-symbol read vector} of a codeword $\mathbf{c}= (c_0,\ldots,c_{n-1}) \in \cc$ in a {\it $b$-symbol read channel} is
\[
\pi_b(\mathbf{c})=\left((c_0,\ldots,c_{b-1}),\ldots,(c_{n-1},c_0,\ldots,c_{b-2}) \right).
\]
Each $\mathbf{c} \in \cc$ has a unique $b$-symbol representation in $(\Ff_q^b)^n$. To characterize the $b$-symbol error-correcting capability, we define the {\it $b$-symbol distance} between any two codewords $\mathbf{a} = (a_0,\ldots,a_{n-1})$ and $\mathbf{c} = (c_0,\ldots,c_{n-1})$ in $\cc$ as
\begin{equation*}
d_b(\mathbf{a},\mathbf{c})=\left| \left \{0\leq i \leq n-1 : (a_i,\ldots, a_{i+b-1}) \neq(c_i,\ldots,c_{i+b-1})\right\}\right|,
\end{equation*}
with subscripts taken modulo $n$. An argument similar to the one in \cite[Section II.A]{Cassuto2011} can be made to confirm that the code $\cc$ equipped with the $b$-symbol distance forms a metric space. The {\it $b$-symbol weight} ${\rm wt}_{b}(\mathbf{c})$ of $\mathbf{c}$ is $d_{b}(\mathbf{c},{\bf 0})$. In the said metric space, the parameters of a $q$-ary linear code $\cc$ is denoted by $(n,k,d_b)^b_q$, with $d_b$ being the minimum $b$-symbol distance of $\cc$. The {\it $b$-symbol weight enumerator} of the linear code $\cc$ is $W_{b}(x) = 1+A_1^bx+ \cdots +A_n^bx^{n}$. The minimum $b$-symbol distance $d_b$ reduces to the minimum Hamming distance $d_{\rm H}$ when $b=1$. We use the notation ${\rm wt}_{\rm H}(\mathbf{c})$ and retain $d_{\rm H}:=d_{\rm H}(\cc)$ for the \emph{Hamming weight} of $\mathbf{c}$ and the \emph{minimum Hamming distance} of $\cc$, respectively.

\subsection{Known Results}

Research on codes in the $b$-symbol metric considers their bounds, constructions, and decoding algorithms. We say that a code \emph{attains} or \emph{meets} a bound if the parameters of the code reach the bound with equality. The $b$-symbol sphere-covering bounds were investigated in \cite{Cassuto2011} and by Chen and Liu in \cite{Chen2023}. Elishco, Gabrys, and Yaakobi in \cite{Elishco2020} established the $b$-symbol linear programming bound. The $b$-symbol Johnson bound was presented in \cite{Elishco2020} for binary codes and extended to nonbinary codes by Chee, Ji, Kiah, Wang, and Jin in \cite{Chen2023}. In the same work, they proposed the $2$-symbol Singleton bound, which was generalized to the $b$-symbol metric by Ding, Zhang, and Ge in \cite{Ding2018}. Based on the $b$-symbol Singleton bound, finding explicit constructions of optimal codes is significant, both theoretically and practically. The attention that the problem has attracted is exemplified by the respective works in \cite{Chee2013,Ding2018,Kai2015,Chen2017,Dinh2018,Liu2023,Luo2023}. 

Cyclic codes are essential in the constructions of codes in the $b$-symbol metric. Yaakobi, Bruck, and Siegel in \cite{Yaakobi2016} proved that any $(n,k,d_2)^2_2$ cyclic code satisfies $d_2 \geq \left\lceil 3 \, d_{\rm H}/2 \right\rceil$, connecting the minimum Hamming distance $d_{\rm H}$ and the $2$-symbol distance $d_2$. Later in \cite{Shi2023}, this relationship was generalized to $d_b \geq \sum_{i=0}^{b-1}\left\lceil d_{\rm H} /q^i\right\rceil$ for any $(n,k,d_b)^b_q$ cyclic code. Shi, \"{O}zbudak, and Sol\'{e} in \cite{Shi2021} developed a geometric approach to cyclic codes in the $b$-symbol metric. They established a relationship between the minimum $(b-1)$-symbol distance and the $b$-symbol distance and proposed a technique to calculate the $b$-symbol weight enumerators of cyclic codes. Using their approach, the $b$-symbol weight enumerators of some irreducible cyclic codes had been subsequently determined in \cite{Zhu2022,Vega2023}.

Given an $[n,k,d_{\rm H}]_q$ code, the Griesmer bound in the Hamming metric, as formulated in \cite{Griesmer1960}, reads $n \geq \sum_{i=0}^{k-1}\left\lceil d_{\rm H} / q^i \right\rceil$. The bound provides a tighter constraint than the Singleton bound, particularly when the alphabet size $q$ is small and the length is sufficiently large. The two bounds, for instance, tells us that no $[n>4,k,d_{\rm H}]_2$ code can attain the Singleton bound, whereas the $[2^k-1,k,2^{k-1}]_2$ simplex code attains the Griesmer bound for any $k \geq 2$. To assess the performance of $q$-ary linear codes in the $b$-symbol metric over small values of $q$, especially for $q=2$, it is imperative to establish a bound in terms of $q$. Very recently, Shi, Zhu, and Helleseth in \cite{Shi2023} formulated a conjecture on the Griesmer bound for cyclic codes in the $b$-symbol metric based on the relationship between the minimum Hamming distance and the $b$-symbol distance. The conjecture is somewhat limited as it exclusively applies to cyclic codes whose dimensions are multiples of $b$.

\subsection{Our Contributions}

This paper focuses on the Griesmer bound for any linear code in the $b$-symbol metric. The contributions can be summarized as follows.

\begin{enumerate}
\item The Griesmer bound of the linear code $\cc$ in the Hamming metric was determined by the {\it residual code}, which is constructed by puncturing $\cc$. This approach, however, does not work when we consider the $b$-symbol distance of $\cc$. Determining the minimum $b$-symbol distance of its residual code is highly challenging. 

We use concatenated codes to convert the $b$-symbol distance to the Hamming distance. We exhibit that each $(n,k,d_b)^b_q$ code gives rise to a $\left[\frac{(q^b-1) \, n}{q-1},k,q^{b-1} \, d_b\right]_q$ code. We use the Griesmer bound in the Hamming metric to derive the Griesmer bound in the $b$-symbol metric, which we formulate in \eqref{Griesmer}. Two well-known codes in \cite[Theorems 2.7 and 3.17]{Shi2023} illustrate that our $b$-symbol Griesmer bound is tight. Our $b$-symbol Griesmer bound in \eqref{Griesmer} takes a different form from the conjecture in \cite{Shi2023} because our bound holds for any linear code, instead of only for cyclic codes.

\item We construct optimal or distance-optimal linear codes in the $b$-symbol metric. We use an optimal or a distance-optimal linear code of relatively short length as an ingredient in a code concatenation technique to build optimal or distance-optimal linear codes of larger lengths with respect to our $b$-symbol Griesmer bound. In Section \ref{sec:4}, we propose two families of distance-optimal codes.
\end{enumerate}

This paper is organized as follows. The next section collects basic results about linear codes and additive characters over finite fields. In Section \ref{sec:griesmer}, we present the Griesmer bound for linear codes in the $b$-symbol metric. In Section \ref{sec:4}, we construct two families of distance-optimal linear codes with respect to the $b$-symbol Griesmer bound. Section \ref{sec:5} concludes this paper.

\section{Preliminaries}\label{sec:pre}

Given two positive integers $s$ and $m$ with $s<m$, we denote by $[m]$ and $[s,m]$ the respective sets $\{1,2,\ldots,m\}$ and $\{s,s+1,\ldots,m\}$. The {\it trace function} from $\Ff_{q^m}$ to $\Ff_q$ is
\[
\Tr_{q^m/q}(x)=x+x^q+\cdots+x^{q^{m-1}} \mbox{, for any } x \in\Ff_{q^m}.
\]
Let $\{\alpha_1,\ldots,\alpha_k\}$ be a {\it basis} of $\Ff_{q^k}$ over $\Ff_q$. From \cite[Page 58]{Lidl1997}, we know of the existence of the {\it dual basis} $\{\beta_1,\ldots,\beta_k\}$ of $\Ff_{q^k}$ over $\Ff_q$ that satisfies, for each $i,j\in[k]$,
\[
\Tr_{q^k/q}(\alpha_i\beta_j) = 
\begin{cases}
1,& \mbox{if } i=j,\\
0,& \mbox{if } i \neq j.
\end{cases}
\]
Given any $g,h\in \Ff_{q^k}$, we write $g=\sum_{i=1}^k g_i \, \alpha_i$ and $h=\sum_{j=1}^k h_j \, \beta_j$ with $g_i,h_j\in\Ff_q$ for each $i, j \in[k]$. We can then identify $g$ and $h$ as vectors $\mathbf{g}=(g_1,\ldots,g_k)$ and $\mathbf{h}=(h_1,\ldots,h_k)$ in $\Ff_q^k$. The {\it Euclidean inner product} of $\mathbf{g}$ and $\mathbf{h}$ is $\mathbf{g}\cdot \mathbf{h}^{\top}=\sum_{i=1}^kg_ih_i$. By definition, $\Tr_{q^m/q}(g \, h)=\mathbf{g}\cdot \mathbf{h}^{\top}$. If $\cc$ is an $[n,k]_q$ linear code with generator matrix $G=\begin{pmatrix}\mathbf{u}_0,\ldots,\mathbf{u}_{n-1}\end{pmatrix}$, then 
\begin{align}\label{Trace-repre}
\cc &=\left\{(\mathbf{h} \cdot \mathbf{u}_0^{\top}, \ldots, \mathbf{h}\cdot \mathbf{u}_{n-1}^{\top}):\mathbf{h}\in\Ff_q^k\right\} \notag \\
&=\left\{(\Tr_{q^k/q}(h \, u_0),\ldots,\Tr_{q^k/q}(h \, u_{n-1})):h\in\Ff_{q^k}\right\}.
\end{align}
Any linear code can be represented by using the trace function in the form of \eqref{Trace-repre}.

The class of {\it simplex codes} is famous among linear codes for meeting the Griesmer bound. Let $\mathcal{S}_{q,b}$ be the $\left[\frac{q^b-1}{q-1},b,q^{b-1}\right]_q$ simplex code with generator matrix $G_{q,b}$ whose columns consist of
\begin{equation}\label{simplex}
\{(1,a_2,\ldots,a_{b-1}):a_2,\ldots,a_{b-1}\in\Ff_q\} \cup\{(0,1,a_3, \ldots,a_{b-1}): a_3,\ldots,a_{b-1}\in\Ff_q\} \cup \ldots \cup \{(0,0,\ldots,0,1)\}.
\end{equation}

Let $\xi_p=e^{\frac{2\pi\sqrt{-1}}{p}}$ be the primitive $p^{\rm th}$ root of complex unity. For each $x\in\Ff_{q}$, the corresponding {\it additive character} of the additive group of $\Ff_{q}$ is
\[
\chi_x(y)=\xi_p^{\Tr_{q/p}(xy)},\ y\in\Ff_q.
\]
The {\it orthogonal relation} of additive characters states
\begin{equation}\label{orthogonal}
\sum_{y\in \Ff_{q}}\chi_x(y) = 
\begin{cases}
0,& \mbox{if } x\neq 0,\\
q,& \mbox{if } x=0.
\end{cases}
\end{equation}

\section{Griesmer Bound for Codes in the $b$-Symbol Metric}\label{sec:griesmer}

Shi, Zhu, and Helleseth made the following conjecture on the Griesmer bound for \emph{cyclic codes} in the $b$-symbol metric.

\begin{conjecture}{\rm \cite[Conjecture 1]{Shi2023}}\label{conjecture}
Let $k$ be a positive integer with $k=t \, b$ for some positive integer $t$. If $\cc$ is an $(n,k,d_b)^b_q$ cyclic code, then $n\geq\sum_{i=0}^{t-1}\left\lceil\frac{d_b}{q^{bi}}\right\rceil$.
\end{conjecture}
Shi, Zhu and Helleseth indicated that the $(n,b,n)^b_q$ cyclic code constructed in \cite[Theorem 3.17]{Shi2023} attains the stated bound. The Griesmer bound in Conjecture \ref{conjecture} is somewhat restricted, as the code $\cc$ is cyclic and the dimension must be a multiple of $b$. Here we provide the Griesmer bound for \emph{any} linear code.

\begin{thm}[$b$-symbol Griesmer Bound]\label{thm-griesmer}
If $\cc$ is any $(n,k,d_b)^b_q$ linear code, then
\begin{equation}\label{Griesmer}
\frac{(q^b-1)n}{q-1}\geq\sum_{i=0}^{k-1}\left\lceil\frac{q^{b-1}d_b}{q^i}\right\rceil.
\end{equation}
\end{thm}
\begin{IEEEproof}
Let $\mathcal{S}_{q,b}$ be a $\left[\frac{q^b-1}{q-1},b,q^{b-1}\right]_q$ simplex code with generator matrix $G_{q,b}$. Let $\phi: \Ff_q^b\rightarrow \mathcal{S}_{q,b}$ be an $\Ff_q$-linear bijection sending $\mathbf{a}\mapsto \mathbf{a}\cdot G_{q,b}$. Let $\mathbf{c}=(c_0,\ldots,c_{n-1})$ be a codeword in $\cc$ and let $\overline{\mathbf{c}}_i=(c_i,\ldots,c_{i+b-1})$, with subscripts taken modulo $n$, for each $i\in[0,n-1]$. The $b$-symbol read vector of $\mathbf{c}$ is $\pi_b(\mathbf{c})=\left(\overline{\mathbf{c}}_0,\ldots,\overline{\mathbf{c}}_{n-1} \right)$. We define the concatenated code $\mathcal{E}$ as
\begin{equation}\label{thm-1-eq1}
\mathcal{E}(\cc)=\left\{\left(\phi(\overline{\mathbf{c}}_0),\ldots,\phi(\overline{\mathbf{c}}_{n-1})\right): \mathbf{c} \in \cc\right\}.
\end{equation}
Since $\phi$ is an $\Ff_q$-linear bijection, $\mathcal{E}(\cc)$ is a linear code with length $\frac{(q^b-1)n}{q-1}$ and dimension $k$. The simplex code $\mathcal{S}_{q,b}$ is a constant-weight code. Given a codeword $\mathbf{c} \in \cc$ with $b$-symbol weight ${\rm wt}_{b}(\mathbf{c})$, the Hamming weight of $\left(\phi(\overline{\mathbf{c}}_0),\ldots,\phi(\overline{\mathbf{c}}_{n-1})\right)$ is $q^{b-1} \, {\rm wt}_{b}(\mathbf{c})$. We can now confirm that $\mathcal{E}(\cc)$ has parameters $\left[\frac{(q^b-1)n}{q-1},k,q^{b-1} \, d_b\right]_q$. By the Griesmer bound in the Hamming metric, we have
\[
\frac{(q^b-1)n}{q-1}\geq\sum_{i=0}^{k-1}\left\lceil\frac{q^{b-1}d_b}{q^i}\right\rceil.
\]
\end{IEEEproof}

In \cite[Theorems 2.7 and 3.17]{Shi2023}, Shi, Zhu and Helleseth proposed two constructions of cyclic codes, with respective parameters $\left(\frac{q^k-1}{\Delta},k,\frac{q^k-q^{k-b}}{\Delta}\right)^b_q$ and $(n,b,n)^b_q$, where $\Delta$ divides $(q-1)$ and $\gcd\left(\Delta,\frac{q^k-1}{\Delta}\right) =1$ . It is easy to check that these cyclic codes attain the Griesmer bound in \eqref{Griesmer}. That is to say, the Griesmer bound in \eqref{Griesmer} is tight. A linear code in the $b$-symbol metric is {\it optimal} if it meets the Griesmer bound in \eqref{Griesmer}. The $(n,k,d_b)^b_q$ linear code is {\it distance-optimal} if there is no $(n,k,d_b+1)^b_q$ linear code.

The concatenated code in \eqref{thm-1-eq1} converts $d_b$ to $d_{\rm H}$. Since the simplex code $\mathcal{S}_{q,b}$ is a constant-weight code, the concatenated code $\mathcal{E}(\cc)$ has
Hamming weight enumerator 
\[
W_{\rm H}(x) = 1+ A_1^b \, x^{q^{b-1}} + A_2^b \, x^{2q^{b-1}} + \cdots + A_n^b \, x^{nq^{b-1}}
\]
if and only if the linear code $\cc$ has $b$-symbol weight enumerator 
\[
W_b(x) = 1 + A_1^b \, x + A_2^b \, x^2 +  \cdots + A_n^b \, x^{n}.
\]
Our next result designs a generator matrix for $\mathcal{E}(\cc)$ in \eqref{thm-2-eq2}, leading to the representation of $\mathcal{E}(\cc)$ in terms of the codewords of $\cc$.

\begin{thm}\label{Hb-relation}
Let $\cc$ be an $(n,k,d_b)^b_q$ code and let $\mathcal{E}(\cc)$ be the concatenated code constructed as in \eqref{thm-1-eq1}. Let $\mathbf{c}=(c_0,\ldots,c_{n-1})\in\cc$. For each $j\in[0,b-1]$, let $\mathbf{c}^j=(c_{j},c_{j+1},\ldots,c_{j-1})$ be its first $j$ cyclic shifts, with subscripts taken modulo $n$. Let $\sigma$ be a permutation on $\left[0,\frac{(q^b-1)n}{q-1}-1\right]$ such that $\sigma\left(\frac{(q^b-1)u}{q-1}+v\right) = v \, n+u$, for each $u\in[0,n-1]$ and $v \in \left[0,\frac{(q^b-1)}{q-1}-1\right]$. Given a codeword $\mathbf{u}=\left(u_0,\ldots,u_{\frac{(q^b-1)n}{q-1}-1}\right)\in\mathcal{E}(\cc)$ formed by $\mathbf{c}$, we have
\begin{align}
\sigma(\mathbf{u})&=\left(u_{\sigma(0)},u_{\sigma(1)},\ldots,u_{\sigma\left(\frac{(q^b-1)n}{q-1}-1\right)}\right)\notag \\
&=\biggl(\left(\mathbf{c}^0+a_1 \, \mathbf{c}^1+ \ldots+ a_{b-1} \, \mathbf{c}^{b-1}\right)_{a_1,\ldots,a_{b-1}\in \Ff_q},  \left(\mathbf{c}^1+a_2 \, \mathbf{c}^2+\ldots+ a_{b-1} \, \mathbf{c}^{b-1}\right)_{a_2,\ldots,a_{b-1}\in \Ff_q}, \notag \\
& \qquad \ldots,\left(\mathbf{c}^{b-2}+ a_{b-1} \, \mathbf{c}^{b-1}\right)_{a_{b-1}\in \Ff_q}, \mathbf{c}^{b-1}\biggr).\label{Hb-representation}
\end{align}
\end{thm}

\begin{IEEEproof}
Let $\begin{pmatrix}\mathbf{g}_0^{\top},\mathbf{g}_1^{\top},\ldots,\mathbf{g}_{n-1}^{\top}\end{pmatrix}$ be a generator matrix of $\cc$. We use its columns, with subscripts taken modulo $n$, to define
\[
G_j :=\begin{pmatrix}\mathbf{g}_{j}^{\top},\mathbf{g}_{j+1}^{\top},\ldots,\mathbf{g}_{j-1}^{\top}\end{pmatrix} \mbox{ and } U_i :=\begin{pmatrix}\mathbf{g}_i^{\top},\ldots,\mathbf{g}_{i+b-1}^{\top}\end{pmatrix},
\]
for any $i,j\in[0,n-1]$. Let $G_{q,b}=\begin{pmatrix}\mathbf{r}_0^{\top}, \ldots, \mathbf{r}_{b-1}^{\top} \end{pmatrix}^{\top}$, written according to \eqref{simplex}, be a generator matrix of the simplex code $\mathcal{S}_{q,b}$. The concatenated code $\mathcal{E}(\cc)$, as defined in \eqref{thm-1-eq1}, can then be expressed as
\begin{equation}\label{thm-2-eq1}
\left\{\left(\mathbf{a}\cdot U_0 \cdot G_{q,b},\ldots,\mathbf{a}\cdot U_{n-1} \cdot G_{q,b}\right): \mathbf{a} \in \Ff_q^k\right\}.
\end{equation}
Bringing the permutation $\sigma$ to the mix, we use \eqref{thm-2-eq1} to express $\sigma\left(\mathcal{E}(\cc)\right)$ as
\begin{equation}\label{thm-2-eq2}
\Big\{(\underbrace{\mathbf{a},\ldots,\mathbf{a}}_b) \cdot T : \mathbf{a} \in \Ff_q^k\Big\}\mbox{, with } T = \begin{pmatrix}\mathbf{r}_0\otimes G_0\\ \vdots\\ \mathbf{r}_{b-1}\otimes G_{b-1} \end{pmatrix}.
\end{equation}
By \eqref{simplex} and \eqref{thm-2-eq2}, given a codeword $\sigma(\mathbf{u})=\left(\mathbf{v}_1,\ldots,\mathbf{v}_{\frac{q^b-1}{q-1}}\right)$ of $\sigma\left(\mathcal{E}(\cc)\right)$, we get
\[
\mathbf{v}_i=
\begin{cases}
\mathbf{c}^0+ a_{i,1} \, \mathbf{c}^1+\cdots+a_{i,b-1} \, \mathbf{c}^{b-1}, & \mbox{ if } i\in[q^{b-1}],\\
\mathbf{c}^1+ a_{i,2} \, \mathbf{c}^2+\cdots+ a_{i,b-1} \, \mathbf{c}^{b-1}, & \mbox{if } i\in[q^{b-1}+1,q^{b-1}+q^{b-2}],\\
\qquad \qquad  \vdots & \qquad \qquad \vdots\\
\mathbf{c}^{b-2}+a_{i,b-1} \, \mathbf{c}^{b-1}, & \mbox{if } i\in\left[\frac{q^b-q^2}{q-1}, \frac{q^b-q^2}{q-1}+q\right],\\
\mathbf{c}^{b-1}, & \hbox{if } i=\frac{q^b-1}{q-1},
\end{cases}
\]
for some $a_{i,j}\in\Ff_q$. When each $a_{i,j}$ in the above system of equation has traversed all elements of $\Ff_q$, we get the representation of $\sigma(\mathbf{u})$ in \eqref{Hb-representation}.
\end{IEEEproof}

\begin{example}
Let us consider the $(4,2,3)_2^2$ code 
\[
\cc=\{(1,1,0,0),(1,0,1,1),(0,1,1,1),(0,0,0,0)\}.
\]
If $\mathcal{S}_{q,b}$ is the $[3,2,2]_2$ simplex code generated by
$
\begin{pmatrix}
1 & 1 & 0\\
0 & 1 & 1
\end{pmatrix}
$, then the concatenated code is
\begin{multline*}
\mathcal{E}(\cc)=\{(1,1,0,1,0,1,0,0,0,0,1,1),(1,0,1,0,1,1,1,1,0,1,1,0),\\
(0,1,1,1,1,0,1,1,0,1,0,1),(0,0,0,0,0,0,0,0,0,0,0,0)\}.
\end{multline*}
Let $\sigma$ be a permutation on $[0,11]$ such that $\sigma(3u+v)=4v+u$ for each $u\in[0,3]$ and $v\in[0,2]$. By Theorem \ref{Hb-relation}, $\mathcal{E}(\cc)$ is permutation equivalent to
\begin{align*}
\sigma\left(\mathcal{E}(\cc)\right)&=\left\{(\mathbf{c}^0,\mathbf{c}^0+\mathbf{c}^1,\mathbf{c}^1)\right\} \\
&=\{(1,1,0,0,1,0,0,1,0,1,0,1),(1,0,1,1,0,1,1,1,1,1,0,0),\\
& \qquad (0,1,1,1,1,0,0,1,1,1,1,0),(0,0,0,0,0,0,0,0,0,0,0,0)
\}.
\end{align*}
\end{example}

\begin{remark}
Let $\mathbf{u}\in\mathcal{E}(\cc)$ be formed by $\mathbf{c}\in\cc$. By \eqref{Hb-representation} and the proof of Theorem \ref{Griesmer}, we have
\begin{equation}\label{b-weight}
{\rm wt}_{\rm H}(\mathbf{u}) = q^{b-1} \, {\rm wt}_{b}(\mathbf{c})=\frac{1}{q-1}\sum_{a_0,\ldots,a_{b-1}\in\Ff_q}{\rm wt}_{\rm H}(a_0 \, \mathbf{c}^0 + a_1 \, \mathbf{c}^1 + \cdots + a_{b-1} \, \mathbf{c}^{b-1}).
\end{equation}
The above relation between the Hamming weight and $b$-symbol weight was established by a combinatorial approach in \cite[Theorem 3.1]{Shi2023}. Theorem \ref{Hb-relation} supplies another proof of this relation by a coding theoretic approach.
\end{remark}

We now discuss a general construction method for codes of large lengths with good minimum $2$-symbol distances from linear codes of shorter lengths with good minimum $2$-symbol distances. We start with an observation on the $b$-symbol weight.

\begin{lem}\label{lem2-2}
Let $\mathbf{u}=(u_0,\ldots,u_{n-1}), \mathbf{v}=(v_0,\ldots,v_{n-1})\in\Ff_q^n$. If $n\geq b$ and $u_i=v_i$ for each $i\in[0,b-2]$, then 
\[
{\rm wt}_{b}((\mathbf{u},\mathbf{v})) = {\rm wt}_{b}(\mathbf{u}) + {\rm wt}_{b}(\mathbf{v}).
\]
\end{lem}
\begin{IEEEproof}
The $b$-symbol read vector of $(\mathbf{u},\mathbf{v})$ is
\begin{multline*}
\pi_b((\mathbf{u},\mathbf{v}))=((u_0,\ldots,u_{b-1}),\ldots,(u_{n-1},v_0,\ldots,v_{b-2}), \\
(v_0,\ldots,v_{b-1}),\ldots,(v_{n-2},v_{n-1},u_0,\ldots,u_{b-2}),(v_{n-1},u_0,\ldots,u_{b-2})).
\end{multline*}
Since $u_i=v_i$ for each $i\in[0,b-2]$, we have
\begin{align*}
\pi_b((\mathbf{u},\mathbf{v}))&=((u_0,\ldots,u_{b-1}),\ldots,(u_{n-1},u_0,\ldots,u_{b-2}),
(v_0,\ldots,v_{b-1}),\ldots,(v_{n-1},v_0,\ldots,v_{b-2}))\\
&=(\pi_b(\mathbf{u}),\pi_b(\mathbf{v})),
\end{align*}
which leads directly to the desired conclusion.
\end{IEEEproof}

\begin{lem}\label{lem2-1}
If $s$ and $k$ are positive integers such that $\gcd(k,q-1)=1$ and $k\geq b$, then there exists an $\left(s\frac{q^k-1}{q-1},k,s\frac{q^k-q^{k-b}}{q-1}\right)^b_q$ linear code.
\end{lem}
\begin{IEEEproof}
Let $n=\frac{q^k-1}{q-1}$ and let $\gamma$ be a primitive element of $\Ff_{q^k}$. We define a cyclic code $\cc$ of length $\frac{q^k-1}{q-1}$ as
\[
\cc=\left\{\mathbf{c}_g=\left(\Tr_{q^k/q}\left(g\gamma^{i(q-1)}\right)\right)_{i=0}^{n-1}:g\in\Ff_{q^k}\right\}.
\]
By \cite[Theorems 2.7]{Shi2023}, $\cc$ has parameters $\left(\frac{q^k-1}{q-1},k,\frac{q^k-q^{k-b}}{q-1}\right)^b_q$ and ${\rm wt}_{b}(\mathbf{c}_g)=\frac{q^k-q^{k-b}}{q-1}$, for each $\mathbf{c}_g\in \cc$. Repeating each codeword of $\cc$ by $s$ times, we define the code $\widetilde{\cc}$ of length $s\frac{q^k-1}{q-1}$ as
\[
\widetilde{\cc}=\left\{\left(\mathbf{c}_g,\ldots,\mathbf{c}_g\right):g\in\Ff_{q^k}\right\}.
\]
By Lemma \ref{lem2-2}, $\widetilde{\cc}$ is an $\left(s\frac{q^k-1}{q-1},k,s\frac{q^k-q^{k-b}}{q-1}\right)^b_q$ cyclic code with constant $b$-symbol weight $s\frac{q^k-q^{k-b}}{q-1}$.
\end{IEEEproof}

We use Lemma \ref{lem2-1} to construct linear codes of large lengths with good minimum $2$-symbol distances.

\begin{thm}\label{thm2-2}
Let $s$ and $k$ be positive integers such that $\gcd(k,q-1)=1$ and $k\geq 2$. If there exists an $\left(m,k,d_2\right)^2_q$ code $\mathcal{D}$, then there exists an $\left(s\frac{q^k-1}{q-1}+m,k,s\frac{q^k-q^{k-2}}{q-1}+d_2\right)^2_q$ code $\overline{\cc}$. Furthermore, if $\mathcal{D}$ is (distance-) optimal with respect to the Griesmer bound in \eqref{Griesmer}, then $\overline{\cc}$ is also (distance-) optimal with respect to the bound.
\end{thm}
\begin{IEEEproof}
We retain the symbols from the proof of Lemma \ref{lem2-1}. Based on \eqref{Trace-repre}, we express $\mathcal{D}$ as
\[
\mathcal{D}=\left\{\mathbf{t}_g=\left(\Tr_{q^k/q}\left(g \, t_i\right)\right)_{i=0}^{m-1}:g\in\Ff_{q^k}\right\},
\]
where $t_i\in\Ff_{q^k}^*=\Ff_{q^k}\setminus\{0\}$ for each $i\in[0,m-1]$. Since $\gcd(k,q-1)=1$, we have $\gcd(n,q-1)=1$, which implies that 
\[
\Ff_{q^k}^*=\left\{\gamma^{in+j(q-1)} : i\in[0,q-2],j\in[0,n-1]\right\}.
\]
Letting $t_0=\gamma^{i_0 \, n + j_0 \, (q-1)}$, we rewrite $\cc$ as
\[
\cc=\left\{\mathbf{c}_g=\left(\Tr_{q^k/q}\left(g \, \gamma^{(j_0+i)(q-1)}\right)\right)_{i=0}^{n-1}:g\in\Ff_{q^k}\right\}.
\]
Let $\mathbf{c}_g=(c_{g,0},\ldots,c_{g,n-1})$ and $\mathbf{t}_g=(t_{g,0},\ldots,t_{g,n-1})$. It follows from $\gamma^{i_0n}\in\Ff_q$ that $c_{g,0}=\gamma^{-i_0n} \, t_{g,0}$. We define the code $\overline{\cc}$ of length $sn+m$ and dimension $k$ as
\[
\overline{\cc}=\Big\{\Big(\underbrace{\mathbf{c}_g,\ldots,\mathbf{c}_g}_s,\gamma^{-i_0n} \, \mathbf{t}_g\Big):g\in\Ff_{q^k}\Big\} \mbox{, with } \gamma^{-i_0n} \, \mathbf{t}_g=(\gamma^{-i_0n} \, t_{g,0},\ldots,\gamma^{-i_0n} \, t_{g,n-1}).
\]
Given a codeword $\overline{\mathbf{c}}\in\overline{\cc}$, it follows from $c_{g,0}=\gamma^{-i_0n} \, t_{g,0}$ and Lemma \ref{lem2-2} that
\begin{align*}
{\rm wt}_2(\overline{\mathbf{c}})&={\rm wt}_2\Big(\underbrace{\mathbf{c}_g,\ldots,\mathbf{c}_g}_s \Big)+{\rm wt}_2(\gamma^{-i_0n} \, \mathbf{t}_g)\\
&={\rm wt}_2 \Big(\underbrace{\mathbf{c}_g,\ldots,\mathbf{c}_g}_s \Big) + {\rm wt}_2 (\mathbf{t}_g).
\end{align*}
Since ${\rm wt}_2(\underbrace{\mathbf{c}_g,\ldots,\mathbf{c}_g}_s)= s \, \frac{q^k-q^{k-2}}{q-1}$ for each $g\in\Ff_{q^k}^*$, the minimum $2$-symbol distance of $\overline{\cc}$ is $s\frac{q^k-q^{k-2}}{q-1}+d_2$. By the Griesmer bound in \eqref{Griesmer}, 
\begin{align}
(q+1) \left(s \, \frac{q^k-1}{q-1}+m\right) & \geq \sum_{i=0}^{k-1} \left\lceil\frac{q\left(s \, \frac{q^k-q^{k-2}}{q-1}+d_2\right)}{q^i}\right\rceil \notag \\
& = \sum_{i=0}^{k-1}\left\lceil \frac{s \, q^{k-1}(q+1)}{q^i}+ \frac{q \, d_2}{q^i}\right\rceil \notag \\
& = (q+1) \, s \, \frac{q^k-1}{q-1}+\sum_{i=0}^{k-1}\left\lceil\frac{q \, d_2}{q^i}\right\rceil.
\end{align}
This inequality is equivalent to 
\begin{equation}\label{eq:equiv1}
(q+1) \, m \geq \sum_{i=0}^{k-1} \left\lceil\frac{q \, d_2}{q^i}\right\rceil,
\end{equation}
confirming that $\overline{\cc}$ is (distance-) optimal with respect to the Griesmer bound in \eqref{Griesmer} if the code $\mathcal{D}$ is (distance-) optimal with respect to the same bound.
\end{IEEEproof}

\begin{remark}
Given $n$ and $k$, let $d_{b}(n,k,q)$ be the {\bf largest possible} minimum $b$-symbol distance based on the Griesmer bound in \eqref{Griesmer}. To measure how good the parameters of codes equipped with $b$-symbol metric can be, we define the \emph{Griesmer gap} as 
\[
{\rm gap}(n,k,d_b,q)=d_{b}(n,k,q)-d_b.
\]
By the proof of Theorem \ref{thm2-2}, if $\mathcal{D}$ has the Griesmer gap $h={\rm gap}(m,k,d_2,q)$, then the Griesmer gap of $\overline{\cc}$ is also $h$. In a nutshell, Theorem \ref{thm2-2} provides a general method to construct linear codes with larger lengths and good minimum $2$-symbol distance from known shorter codes. For example, to find all optimal or distance-optimal binary linear codes of dimension $4$ with respect to the Griesmer bound in \eqref{Griesmer}, we only need to find all optimal or distance-optimal $(n,4,d_2)_2^2$ linear codes for $1 \leq n \leq 14$. Using Theorem \ref{thm2-2}, we obtain all optimal or distance-optimal $(15s+n,4,12s+d_2)_2^2$ codes with respect to the same bound.
\end{remark}

\section{Constructions of Linear Codes in $b$-Symbol Metric}\label{sec:4}

In this section, we provide two constructions of distance-optimal linear codes in the $b$-symbol metric.

\begin{thm}\label{con1}
Let $k$ be a positive integer with $k \geq b$. Let $\gamma$ be a primitive element of $\Ff_{q^k}$. If $\cc$ is given by
\[
\left\{\mathbf{c}_{g,y}=\left(\Tr_{q^k/q}\left(g \, \gamma^{i}\right)+y\right)_{i=0}^{q^k-2}:g\in\Ff_{q^k} \mbox{ and } y\in\Ff_q\right\},
\]
then $\cc$ has parameters $(q^k-1,k+1,q^k-q^{k-b}-1)^b_q$ and is distance-optimal with respect to the Griesmer bound in \eqref{Griesmer}.
\end{thm}

\begin{IEEEproof}
We use \eqref{b-weight} to determine the Hamming weight of $a_0 \, \mathbf{c}_{g,y}^0 + \cdots + a_{b-1} \, \mathbf{c}_{g,y}^{b-1}$ for each $a_0,\ldots,a_{b-1}\in\Ff_q$. We note that ${\rm wt}_{b}(\mathbf{c}_{0,0})=0$ and consider the case when $(g,y) \neq (0,0)$. Writing $n=q^k-1$, by how $\cc$ is defined, we have
\[
a_0 \, \mathbf{c}_{g,y}^0 + \cdots + a_{b-1} \, \mathbf{c}_{g,y}^{b-1} = \left(\Tr_{q^k/q}\left(g \, \gamma^{i}\left(\sum_{j=0}^{b-1} a_j \, \gamma^{j}\right)\right)+y\sum_{j=0}^{b-1}a_j\right)_{i=0}^{n-1}.
\]
Let $\chi(z)=\xi_p^{\Tr_{q/p}(z)}$ and $\varphi(x)=\xi_p^{\Tr_{q^k/p}(x)}$ be the respective additive characters of $\Ff_q$ and $\Ff_{q^k}$. For each $x\in\Ff_{q^k}$, we have $\varphi(x)=\chi(\Tr_{q^k/q}(x))$. Using the orthogonal relation of additive characters in \eqref{orthogonal}, we get
\begin{align*}
{\rm wt}_{\rm H} \left(a_0 \, \mathbf{c}_{g,y}^0 + \cdots + a_{b-1} \, \mathbf{c}_{g,y}^{b-1}\right) &= n-\frac{1}{q} \sum_{i=0}^{n-1} \sum_{z\in\Ff_q} \chi \left(z
\left(\Tr_{q^k/q} \left(g \, \gamma^{i}\left(\sum_{j=0}^{b-1} a_j \, \gamma^{j}\right)\right)+y\sum_{j=0}^{b-1}a_j\right)\right)\\
&=\frac{q-1}{q}n-\frac{1}{q} \sum_{z\in\Ff_q^*} \sum_{i=0}^{n-1} \varphi\left(z \, g \, \gamma^{i} \left(\sum_{j=0}^{b-1} a_j \, \gamma^{j}\right)\right) \chi \left(z \, y \, \sum_{j=0}^{b-1}a_j\right)\\
&=\frac{q-1}{q} n - \frac{1}{q} \sum_{z\in\Ff_q^*} \chi\left(z \, y \, \sum_{j=0}^{b-1} a_j\right) \sum_{i=0}^{n-1} \varphi\left(z \, g \, \gamma^{i}\left(\sum_{j=0}^{b-1} a_j \, \gamma^{j}\right)\right).
\end{align*}
Since $k\geq b$ and $\gamma$ is a primitive element of $\Ff_{q^k}$, the set $\{1,\gamma,\ldots,\gamma^{b-1}\}$ is linearly independent over $\Ff_q$. If $(a_0,\ldots,a_{b-1}) = \mathbf{0}$, then ${\rm wt}_{\rm H}(a_0 \, \mathbf{c}_{g,y}^0 + \cdots+ a_{b-1} \, \mathbf{c}_{g,y}^{b-1}) = 0$. Next, assuming that $(a_0,\ldots,a_{b-1})\neq \mathbf{0}$, we compute ${\rm wt}_{\rm H}(a_0 \, \mathbf{c}_{g,y}^0+\cdots + a_{b-1} \, \mathbf{c}_{g,y}^{b-1})$ in three cases.

\begin{enumerate}[wide, itemsep=0pt, leftmargin =0pt, widest={{\bf Case $2$}}]
\item[{\bf Case $1$}:] If $g=0$ and $y\neq 0$, then 
\begin{align*}
{\rm wt}_{\rm H} \left(a_0 \, \mathbf{c}_{g,y}^0 + \cdots + a_{b-1} \, \mathbf{c}_{g,y}^{b-1}\right) &=\frac{q-1}{q}n-\frac{n}{q}\sum_{z\in\Ff_q^*}\chi\left(zy\sum_{j=0}^{b-1}a_j\right)\\
&=\begin{cases}
0, & \mbox{if } \sum_{j=0}^{b-1}a_j=0, \\
n, & \mbox{if } \sum_{j=0}^{b-1}a_j\neq 0.
\end{cases}
\end{align*}
\item[{\bf Case $2$}:] If $g\neq 0$ and $y=0$, then 
\begin{align*}
{\rm wt}_{\rm H} \left(a_0 \, \mathbf{c}_{g,y}^0 + \cdots + a_{b-1} \, \mathbf{c}_{g,y}^{b-1}\right) &=\frac{q-1}{q}n-\frac{1}{q}\sum_{z\in\Ff_q^*}\sum_{i=0}^{n-1}\varphi\left(zg\gamma^{i}\left(\sum_{j=0}^{b-1}a_j\gamma^{j}\right)\right)\\
&=\frac{q-1}{q}(n+1).
\end{align*}
\item[{\bf Case $3$}:] If $g\neq 0$ and $y\neq 0$, then 
\begin{align*}
{\rm wt}_{\rm H} \left(a_0 \, \mathbf{c}_{g,y}^0 + \cdots + a_{b-1} \, \mathbf{c}_{g,y}^{b-1}\right) &=\frac{q-1}{q}n+\frac{1}{q}\sum_{z\in\Ff_q^*}\chi\left(zy\sum_{j=0}^{b-1}a_j\right)\\
&=
\begin{cases}
\frac{q-1}{q}(n+1), & \mbox{if } \sum_{j=0}^{b-1}a_j=0, \\
\frac{(q-1)n-1}{q},& \mbox{if } \sum_{j=0}^{b-1}a_j\neq 0.
\end{cases}
\end{align*}
\end{enumerate}
If $a_0=1$ and $a_1=\ldots=a_{b-1}=0$, then ${\rm wt}_{\rm H}(\mathbf{c}_{g,y}^0)={\rm wt}_{\rm H}(\mathbf{c}_{g,y})>0$ whenever $(g,y)\neq (0,0)$. This ensures that the dimension of the code $\cc$ is $k+1$. It is also clear that
\[
\left|\left\{(a_0,\ldots,a_{b-1})\in\Ff_q^b:\sum_{j=0}^{b-1}a_j=0\right\}\right|=q^{b-1}.
\]

Summarizing the above three cases and consulting \eqref{b-weight}, we obtain
\begin{align*}
{\rm wt}_{b}(\mathbf{c})& = \frac{1}{(q-1)q^{b-1}} \left(\sum_{a_0,\ldots, a_{b-1}\in\Ff_q}{\rm wt}_{\rm H} (a_0 \, \mathbf{c}^0 + a_1 \, \mathbf{c}^1 + \cdots + a_{b-1} \, \mathbf{c}^{b-1}) \right)\\
&=\begin{cases}
q^k-1, & \mbox{if } g=0 \mbox{ and } y \neq 0,\\
q^k-q^{k-b},& \mbox{if } g\neq 0 \mbox{ and } y=0,\\
q^k-q^{k-b}-1,& \mbox{if } g\neq 0 \mbox{ and } y\neq 0.
\end{cases}
\end{align*}
We know, therefore, that $\cc$ has parameters $(q^k-1,k+1,q^k-q^{k-b}-1)^b_q$.

For a contradiction, let us assume that there exists a $(q^k-1,k+1,q^k-q^{k-b})^b_q$ code $\mathcal{C}'$ of the same length and dimension as $\cc$ but with $d_b(\mathcal{C}') = d_b(\cc) + 1$. By the Griesmer bound in \eqref{Griesmer}, 
\[
\frac{(q^b-1)(q^k-1)}{q-1}\geq \sum_{i=0}^{k}\left\lceil\frac{q^{b-1}(q^k-q^{k-b})}{q^i}\right\rceil=\frac{(q^b-1)(q^k-1)}{q-1}+q^{b-1},
\]
which is absurd. The nonexistence of such a code $\mathcal{C}'$ ensures that $\cc$ is distance-optimal with respect to the bound.
\end{IEEEproof}

The code $\cc$ constructed in Theorem \ref{con1} is cyclic. Extending the code $\cc$ to the code $\widetilde{\cc}$ of length $q^k$ gives us the following construction.

\begin{thm}\label{con2}
Let $k$ be a positive integer such that $k\geq b$. Let $\gamma$ be a primitive element of $\Ff_{q^k}$. If the code $\widetilde{\cc}$ is defined as
\[
\left\{\widetilde{\mathbf{c}}_{g,y}=\left(\left(\Tr_{q^k/q}\left(g \, \gamma^{i}\right)+y\right)_{i=0}^{q^k-2},y\right):g\in\Ff_{q^k} \mbox{ and } y\in\Ff_q\right\},
\]
then it has parameters $(q^k,k+1,q^k-q^{k-b})^b_q$ and is distance-optimal with respect to the Griesmer bound in \eqref{Griesmer}.
\end{thm}
\begin{IEEEproof}
By Theorem \ref{con1}, $\widetilde{\cc}$ has dimension $k+1$. Let $n=q^k-1$. Given a codeword $\widetilde{\mathbf{c}}_{g,y}\in \widetilde{\cc}$, its $b$-symbol read vector is
\[
\pi_b(\widetilde{\mathbf{c}}_{g,y})=\left((c_i,\ldots, c_{i+b-1})\right)_{i=0}^{n},
\]
with subscripts taken modulo $(n+1)$.
Let $\chi(z)=\xi_p^{\Tr_{q/p}(z)}$ and $\varphi(x)=\xi_p^{\Tr_{q^k/p}(x)}$ be the respective additive characters of $\Ff_q$ and $\Ff_{q^k}$. We suppose that $\sum_{i=j_1}^{j_2} x_i=0$ if $j_2<j_1$, where $x_i$ is a complex number. By the orthogonal relation of additive characters in \eqref{orthogonal}, 
\begin{align*}
q^k-{\rm wt}_{b}(\widetilde{\mathbf{c}}_{g,y})& =\frac{1}{q^b} \sum_{i=0}^n \sum_{a_0\in\Ff_q}\chi(a_0 \, c_i) \cdots \sum_{a_{b-1}\in\Ff_q}\chi(a_{b-1} \, c_{i+b-1})\\
&=\frac{1}{q^b} \sum_{a_0,\ldots,a_{b-1}\in\Ff_q} \sum_{i=0}^n \chi(a_0 \, c_i+ \cdots+ a_{b-1} \, c_{i+b-1})\\
&=\frac{1}{q^b} \sum_{a_0,\ldots, a_{b-1}\in\Ff_q} \Biggl[\sum_{i=0}^{n-b} \varphi \left(g\gamma^{i}\left(\sum_{j=0}^{b-1} a_j \, \gamma^{j}\right)\right) \chi \left(y\sum_{j=0}^{b-1} a_j\right) + \\
& \qquad \sum_{i=n-b+1}^n \varphi \left(g\, \gamma^{i} \left(\sum_{j=0}^{n-i-1} a_j \, \gamma^{j}+ \sum_{j=n-i}^{b-2} a_{j+1} \, \gamma^{j}\right)\right) \chi \left(y\sum_{j=0}^{b-1} a_j\right)\Biggr]\\
&=\frac{1}{q^b} \sum_{a_0,\ldots,a_{b-1}\in\Ff_q} \Biggl[\sum_{i=0}^{n-1} \varphi\left(g \, \gamma^{i} \left(\sum_{j=0}^{b-1} a_j \, \gamma^{j}\right)\right) \chi \left(y\sum_{j=0}^{b-1} a_j\right) - \\
&\qquad \sum_{i=n-b+1}^{n-1} \varphi\left(g\, \gamma^{i} \left(\sum_{j=0}^{b-1} a_j \, \gamma^{j}\right)\right) \chi \left(y\sum_{j=0}^{b-1} a_j\right) +\\
& \qquad \sum_{i=n-b+1}^n \varphi \left(g\, \gamma^{i} \left(\sum_{j=0}^{n-i-1} a_j \, \gamma^{j} + \sum_{j=n-i}^{b-2} a_{j+1} \, \gamma^{j}\right)\right) \chi \left(y\sum_{j=0}^{b-1} a_j\right)\Biggr].
\end{align*}
We write
\begin{align*}
X&=\sum_{a_0,\ldots,a_{b-1}\in\Ff_q}\sum_{i=0}^{n-1} \varphi \left(g \, \gamma^{i}\left(\sum_{j=0}^{b-1} a_j \, \gamma^{j}\right)\right) \chi\left(y\sum_{j=0}^{b-1} a_j\right), \\ 
Y&=\sum_{a_0,\ldots,a_{b-1}\in\Ff_q} \sum_{i=n-b+1}^{n-1} \varphi \left(g \, \gamma^{i} \left(\sum_{j=0}^{b-1} a_j \, \gamma^{j}\right)\right) \chi \left(y\sum_{j=0}^{b-1} a_j\right),\\
Z&=\sum_{a_0,\ldots,a_{b-1}\in\Ff_q} \sum_{i=n-b+1}^n \varphi \left(g \, \gamma^{i}\left(\sum_{j=0}^{n-i-1} a_j \, \gamma^{j}+\sum_{j=n-i}^{b-2} a_{j+1} \, \gamma^{j}\right)\right) \chi\left(y \sum_{j=0}^{b-1} a_j\right).
\end{align*}

Our next task is to compute ${\rm wt}_{b}(\widetilde{\mathbf{c}}_{g,y})$ in three cases.
\begin{enumerate}[wide, itemsep=0pt, leftmargin =0pt, widest={{\bf Case $2$}}]
\item[{\bf Case $1$}:] 
If $g=0$ and $y\neq 0$, then  $\widetilde{\mathbf{c}}_{0,y}=(y,y,\ldots,y)$, which implies ${\rm wt}_{b}(\widetilde{\mathbf{c}}_{0,y})=n+1$.
\item[{\bf Case $2$}:] 
If $g\neq 0$ and $y=0$, then 
\[
\widetilde{\mathbf{c}}_{g,0}=\left(\left(\Tr_{q^k/q} \left(g \, \gamma^{i}\right)\right)_{i=0}^{q^k-2},0\right),
\]
making it easy to infer that 
\[
{\rm wt}_{b}(\widetilde{\mathbf{c}}_{g,0}) \geq {\rm wt}_{b}\left(\left(\Tr_{q^k/q}\left(g \, \gamma^{i}\right)\right)_{i=0}^{q^k-2}\right).
\]
It follows from Case $2$ in the proof of Theorem \ref{con1} that  ${\rm wt}_{b}(\widetilde{\mathbf{c}}_{g,y})\geq q^k-q^{k-b}$.
\item[{\bf Case $3$}:] 
If $g\neq 0$ and $y \neq 0$, then $Z=0$ since $\sum_{a_{n-i}\in\Ff_q} \chi \left(y \, a_{n-i}\right)=0$ for each $i\in[n-b+1,n]$. Using the orthogonal relation of additive characters of $\Ff_q$, we get
\[
X = n-\sum_{\substack{a_0,\ldots,a_{b-1}\in\Ff_q,\\(a_0,\ldots,a_{b-1})\neq \mathbf{0}}} \chi \left(y\sum_{j=0}^{b-1}a_j\right)=n-\left(\sum_{a_0,\ldots,a_{b-1}\in\Ff_q}\chi\left(y\sum_{j=0}^{b-1}a_j\right)-1\right)=n+1.
\]
For each $i\in[n-b+1,n-1]$, we confirm that
\begin{align*}
\sum_{a_0,\ldots,a_{b-1}\in\Ff_q} \varphi\left(g \, \gamma^{i} \left(\sum_{j=0}^{b-1} a_j \, \gamma^{j}\right)\right) \chi \left(y\sum_{j=0}^{b-1} a_j\right)
&=\prod_{j=0}^{b-1} \sum_{a_j\in\Ff_q}\chi\left(a_j \, \Tr_{q^k/q} \left(g \, \gamma^{i+j} + y\right)\right)\\
&=\begin{cases}
q^b, & \mbox{if } \Tr_{q^k/q}\left(g \, \gamma^{i+j} + y\right)=0 \mbox{ for each } j \in[0,b-1], \\
0 ,& \mbox{otherwise.}\\
\end{cases}
\end{align*}
Hence, $Y\geq 0$. Since $q^k-{\rm wt}_{b}(\widetilde{\mathbf{c}}_{g,y})=\frac{1}{q^b}\left(X-Y+Z\right)$, we arrive at 
${\rm wt}_{b}(\widetilde{\mathbf{c}}_{g,y}) \geq q^k-q^{k-b}$.
\end{enumerate}

Summarizing the above three cases, $\widetilde{\cc}$ has parameters $(q^k,k+1,d_b\geq q^k-q^{k-b})^b_q$. If $d_b=q^k-q^{k-b}+1$, then, by the Griesmer bound in \eqref{Griesmer}, 
\[
\frac{(q^b-1)q^k}{q-1}\geq \sum_{i=0}^{k}\left\lceil\frac{q^{b-1}(q^k-q^{k-b}+1)}{q^i}\right\rceil=\frac{(q^b-1)q^k}{q-1}+q^{b-1}+k-b,
\]
which is absurd. Thus, $\widetilde{\cc}$ must be a $(q^k,k+1,q^k-q^{k-b})^b_q$ code which is distance-optimal with respect to the Griesmer bound in \eqref{Griesmer}.
\end{IEEEproof}

\section{Concluding remarks}\label{sec:5}

Theorem \ref{thm-griesmer} establishes the $b$-symbol Griesmer bound by concatenating a simplex code of dimension $b$ and an $(n,k,d_b)^b_q$ linear code. The theorem relates the $b$-symbol distance to the Hamming distance. If the simplex code of dimension $b$ is replaced by an $[N,b,D]_q$ linear code, the concatenated code $\mathcal{E}(\cc)$ has parameters $[N \, n, k, \geq D \, d_b]_q$. The corresponding Griesmer bound takes the form $N \, n \geq \sum_{i=0}^{k} \left\lceil\frac{D \, d_b}{q^i}\right\rceil$, which can be improved by computing the actual minimum Hamming distance of $\mathcal{E}(\cc)$. A particularly interesting direction to investigate is to improve the $b$-symbol Griesmer bound in \eqref{Griesmer} for some special classes of linear codes by evaluating the minimum Hamming distance of the concatenated code $\mathcal{E}(\cc)$.

In Section \ref{sec:4}, we have constructed two families of distance-optimal linear codes with respect to the Griesmer bound in \eqref{Griesmer}. Simplex and punctured simplex codes are well-known for meeting the Griesmer bound in the Hamming metric. They, in fact, are often called Griesmer codes in the Hamming space. We believe that simplex and punctured simplex codes have good $b$-symbol distances. Determining the exact $b$-symbol distances of simplex and punctured simplex codes is an interesting challenge.


\end{document}